\RequirePackage{fix-cm}

\documentclass[referee,natbib,amsmath,amssymb,floatfix,endfloats*]{svjour3}     

\usepackage{graphicx}
\usepackage{dcolumn}
\usepackage{bm}
\usepackage{amssymb,amsfonts,amsmath}
\usepackage{latexsym}
\DeclareMathAlphabet{\mathsfsl}{OT1}{cmr}{bx}{it}
\journalname{Microfluidics and Nanofluidics}

\begin{document}

\title{Molecular dynamics simulations of oscillatory Couette flows with slip boundary conditions}

\titlerunning{Oscillatory Couette flows with slip boundary conditions}     

\author{Nikolai V. Priezjev}

\institute{Nikolai V. Priezjev                  \at
          Department of Mechanical Engineering  \\
          Michigan State University             \\
          East Lansing, Michigan 48824          \\
          \email{priezjev@egr.msu.edu}          }

\date{Received: date / Accepted: date}

\maketitle

\begin{abstract}

The effect of interfacial slip on steady-state and time-periodic
flows of monatomic liquids is investigated using non-equilibrium
molecular dynamics simulations.  The fluid phase is confined between
atomically smooth rigid walls, and the fluid flows are induced by
moving one of the walls.   In steady shear flows, the slip length
increases almost linearly with shear rate.   We found that the
velocity profiles in oscillatory flows are well described by the
Stokes flow solution with the slip length that depends on the local
shear rate.   Interestingly, the rate dependence of the slip length
obtained in steady shear flows is recovered when the slip length in
oscillatory flows is plotted as a function of the local shear rate
magnitude.   For both types of flows, the friction coefficient at
the liquid-solid interface correlates well with the structure of the
first fluid layer near the solid wall.




\end{abstract}

\section{Introduction}
\label{sec:Intro}

The rational design of micro- and nanofluidics devices requires an
accurate prediction of time-dependent flows at the submicron
scales~\cite*[]{Karniadakis05}.   It is well recognized that fluid
flows in confined systems can be significantly affected by slip
boundary conditions.   The velocity discontinuity is usually
quantified via the slip length, which is defined as an extrapolated
distance with respect to the liquid-solid interface to the point
where the relative tangential velocity component vanishes.   At
sufficiently high oscillation frequencies, the fluid slip velocity
might not be in phase with the substrate velocity, and, therefore,
the slip length in general is a complex
number~\cite*[]{Willmott07,NgWang11}. Oscillatory flows with slip
boundary conditions of Newtonian liquids were studied experimentally
using quartz crystal
microbalance~\cite*[]{Ferrante94,EllisHayward03,Du04}.   Molecular
dynamics (MD) simulations are particularly well suited to
investigate the effects of materials properties of liquid-solid
interfaces on flow boundary conditions~\cite*[]{Bocquet07,Li10}.

A direct comparison between MD simulations and continuum analysis of
steady-state flows over chemically
textured~\cite*[]{Priezjev05,Qian05,Priezjev11} or periodically
corrugated~\cite*[]{Priezjev06,Niavarani08} surfaces has indicated
that there is an excellent agreement of the velocity profiles and
the effective slip lengths when the typical length scale of
substrate inhomogeneities is about an order of magnitude larger than
the molecular size.  In the case of time-dependent flows, the main
difficulty in extracting flow properties using MD simulations is
that averaging over thermal fluctuations has to be repeated over
many cycles, which often requires significant computational
recourses.  The MD simulations of oscillatory flows of monatomic and
polymeric fluids have shown that the velocity profiles with no-slip
boundary conditions can be well described by the continuum
mechanics~\cite*[]{Khare01}.  The flow profiles with a finite slip
velocity in oscillatory flows were reported at low fluid densities
and weak wall-fluid interactions~\cite*[]{Hansen06,Hansen07}.  More
recently, it was shown that the slip length depends on the magnitude
and gradient of shear rate near the oscillating wall, and the fluid
slip velocity lags the wall velocity, thus leading to a hysteresis
loop~\cite*[]{Mohseni12}.   However, the dependence of the slip
length on the local shear rate or oscillation frequency and
amplitude has not yet been systematically investigated.

Molecular dynamics simulations by \cite{Nature97} have shown that in
steady shear flow of Netwonian liquids over atomically smooth
crystalline surfaces, the slip length is constant at relatively low
shear rates and it increases nonlinearly at higher rates.   Later,
the nonlinear shear rate dependence of the slip length was
repeatedly observed in MD
studies~\cite*[]{Priezjev04,Yang05,Priezjev07,Asproulis10,Niavarani10,WangZhao11,Freund11,WangZhaoActa11,Kannam12,Priezjev12}.
It was also found that the slip length varies almost linearly with
shear rate when liquid and solid phases form incommensurable
structures at the interface and the wall-fluid interaction energy is
sufficiently high~\cite*[]{Priezjev07,PriezjevJCP}.   More recently,
it was demonstrated that the characteristic slip velocity associated
with the onset of the nonlinear slip regime correlates well with the
diffusion time of fluid monomers over the distance between the
nearest minima of the periodic surface potential at
equilibrium~\cite*[]{Priezjev10}.   One of the motivations of the
present study is to examine whether shear rate dependent slip
boundary conditions observed in steady-state flows are valid for
time-dependent flows.

For steady-state flows, a number of previous MD studies have
established a correlation between the degree of slip and fluid
structure induced by the periodic surface
potential~\cite*[]{Thompson90,Barrat99fd,Priezjev07,Priezjev08,Priezjev09,Priezjev10,Zhang12}.
In particular, it was shown that for atomically smooth, weakly
attractive surfaces, the friction coefficient at the liquid-solid
interface is well described by a function of the product of the main
peak in the static structure factor and the contact density, both
evaluated in the first fluid layer~\cite*[]{Priezjev10}.  However,
the situation is less clear for time-dependent flows where the
surface-induced fluid structure and boundary slip might have a phase
difference (especially at high oscillation frequencies), and the
conclusions obtained for steady-state flows might not be valid.  In
the present study, we performed a comparative analysis of the fluid
structure and the friction coefficient for steady-state and
time-periodic flows.

In this paper, non-equilibrium steady-state and time-periodic
molecular dynamics simulations are performed to investigate Couette
flows with slip boundary conditions. First, the rate dependence of
the slip length is computed in steady shear flows. Then, the
velocity profiles in oscillatory flows are compared with the Stokes
flow solutions in a wide range of frequencies.  We find that the
slip length as a function of the local shear rate estimated at the
stationary and oscillating walls is in good agreement with the
results obtained for steady flows.  We will also show that, for both
types of flows, the friction coefficient at the liquid-solid
interface correlates well with the structure of the fluid layer in
contact with the solid wall.

The rest of the paper proceeds as follows.  In the next section, the
details of molecular dynamics simulations are described.  In
Sect.~\ref{sec:Results}, we briefly review the Stokes flow solution
for oscillatory Couette flows, present the results for steady-state
shear flows, and then analyze the velocity and density profiles,
slip length, and fluid structure in oscillatory flows.  Conclusions
are given in the last section.

\section{Molecular dynamics simulation model}
\label{sec:MD_Model}

\begin{figure}[t]
\includegraphics[width=8.0cm,angle=0]{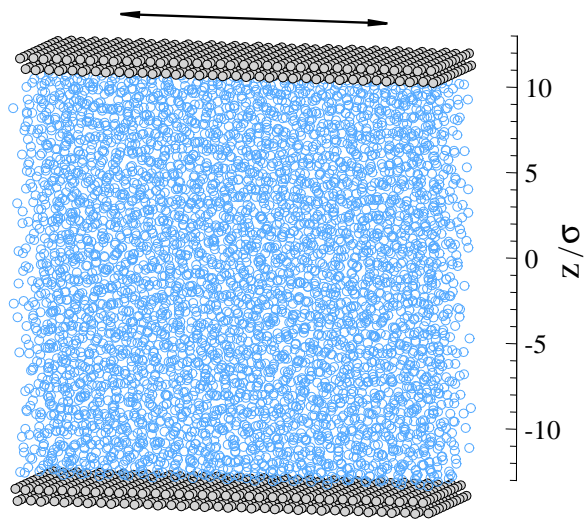}
\caption{(Color online) Positions of fluid monomers (open blue
circles) and wall atoms (filled gray circles). The upper wall
oscillates with the angular frequency $\omega$ in the $\hat{x}$
direction (indicated by the double-sided arrow), while the lower
wall is always stationary. } \label{Fig:snapshot}
\end{figure}

The model system consists of $N_{f}=4608$ fluid monomers confined
between rigid atomistic walls as shown in Fig.\,\ref{Fig:snapshot}.
The pairwise interaction between any two fluid monomers is modeled
by the Lennard-Jones (LJ) potential
\begin{equation}
V_{LJ}(r)=4\,\varepsilon\,\Big[\Big(\frac{\sigma}{r}\Big)^{12}\!-\Big(\frac{\sigma}{r}\Big)^{6}\,\Big],
\label{Eq:LJ}
\end{equation}
where $\varepsilon$ and $\sigma$ are the energy and length scales,
and the cutoff radius $r_c=2.5\,\sigma$.  In our simulations, the
same parameters are used to describe the interaction between fluid
monomers and wall atoms; namely, $\varepsilon_{\rm wf}=\varepsilon$,
$\sigma_{\rm wf}=\sigma$, and $r_c=2.5\,\sigma$. The wall atoms are
fixed rigidly at the lattice sites and do not interact with each
other.

The viscous heating generated in the oscillating flow was removed by
means of the Langevin thermostat, which was coupled only to the
equation of motion perpendicular to the plane of shear as follows:
\begin{eqnarray}
\label{Eq:Langevin_x}
m\ddot{x}_i & = & -\sum_{i \neq j} \frac{\partial V_{ij}}{\partial x_i}, \\
\label{Eq:Langevin_y}
m\ddot{y}_i + m\Gamma\dot{y}_i & = & -\sum_{i \neq j} \frac{\partial V_{ij}}{\partial y_i} + f_i, \\
\label{Eq:Langevin_z}
m\ddot{z}_i & = & -\sum_{i \neq j} \frac{\partial V_{ij}}{\partial z_i}, %
\end{eqnarray}
where $\Gamma=1.0\,\tau^{-1}$ is the friction coefficient and $f_i$
is a random force with zero mean and variance $\langle
f_i(0)f_j(t)\rangle=2mk_BT_{L}\Gamma\delta(t)\delta_{ij}$
\cite*[]{Thompson90}.  In our setup, the temperature of the Langevin
thermostat is set $T_{L}=1.1\,\varepsilon/k_B$, where $k_B$ is the
Boltzmann constant.  The equations of motion were integrated
numerically using the fifth-order Gear predictor-corrector algorithm
\cite*[]{Allen87} with a time step $\triangle t=0.005\,\tau$, where
$\tau=\sqrt{m\sigma^2/\varepsilon}$ is the characteristic LJ time.
The length, energy, and time scales for liquid argon are
$\sigma=0.34\,$nm, $\varepsilon/k_{B}=120\,$K, and
$\tau=2.16\times10^{-12}\,$s \cite*[]{Allen87}.


The fluid phase of density $\rho=0.81\,\sigma^{-3}$ is confined
between two crystalline walls with density
$\rho_w=2.73\,\sigma^{-3}$, as illustrated in
Fig.\,\ref{Fig:snapshot}. Each wall consists of two layers of atoms
arranged rigidly on sites of the face-centered cubic (fcc) lattice.
The lateral dimensions in the $xy$ plane are measured
$L_x=25.03\,\sigma$ and $L_y=9.63\,\sigma$, and the channel width is
fixed $h=23.58\,\sigma$.  Periodic boundary conditions were applied
in the $xy$ plane parallel to the solid walls.  This simulation
setup is very similar to the one used previously for steady
Poiseuille flows \cite*[]{Priezjev07,PriezjevJCP}, except that in
the present study the system size in the $\hat{y}$ direction is
slightly larger. In steady shear flows, the fluid viscosity $\mu =
(2.15\pm0.15)\,\varepsilon\tau\sigma^{-3}$ was found to be shear
rate and temperature independent for $\dot{\gamma}\tau \lesssim
0.072$ and $1.1 \leqslant T\,k_B/\varepsilon \leqslant 1.35$
\cite*[]{Niavarani10}.


To simulate the oscillatory Couette flow, the upper wall velocity
was varied in the $\hat{x}$ direction with the angular frequency
$\omega$ and amplitude $U$, while the lower wall always remained
stationary.   In the present study, the oscillation frequency was
set $\omega\tau=10^{-1}, 10^{-2}, 10^{-3}$, and $10^{-4}$ (see
Table\,\ref{Tab:table_I}). Before the averaging procedure, the
steady-periodic flow was equilibrated during the time interval of
about $5\times10^4\,\tau$. The measurements of the velocity,
density, and temperature profiles were made at discrete times
$\omega t_n = n\,\pi/4 + 2\,\pi\,m$, where $n = 0, 1,...,7$ and $m$
is the integer.   These profiles were averaged within horizontal
bins of thickness $\Delta z=0.01\,\sigma$ during the time interval
$T/100$, where $T=2\,\pi/\omega$ is the period of oscillation.    A
typical simulation time at low shear rates is about
$1.2\times10^6\,\tau$.

\section{Results}
\label{sec:Results}

\subsection{Hydrodynamic predictions}
\label{subsec:Continuum}


The problem of fully-developed oscillatory viscous flow confined
between two parallel walls with slip boundary conditions was
considered analytically by \cite{Vafai04} and \cite{Matthews09}.
Below, we briefly review the problem and its solution for the flow
geometry depicted in Fig.\,\ref{Fig:snapshot}.  The
$\hat{x}$-component of the momentum equation (parallel to the walls)
is given by
\begin{equation}
\rho \frac{\partial u_{x}}{\partial t} = \mu \frac{\partial^2
u_{x}}{\partial z^2}, \label{Eq:NS}
\end{equation}
where $\mu$ and $\rho$ are the fluid viscosity and density.  The
boundary conditions at the top and bottom walls are specified as
follows:
\begin{eqnarray}
\label{Eq:BCtop}
z & = & z_{top}: u_{x} = U \textrm{sin}(\omega t) - L_1 \frac{\partial u_{x}}{\partial z}, \\
\label{Eq:BCbot} z & = & z_{bot}: u_{x} = L_2 \frac{\partial
u_{x}}{\partial z},
\end{eqnarray}
where $U$ is the amplitude and $\omega$ is the frequency of
oscillation. The slip lengths at the upper and lower walls $L_1\neq
L_2$ are assumed to be constant \cite*[]{Matthews09}. We note that
the special case $L_1 = L_2$ was considered by \cite{Vafai04}.


The solution of the problem Eq.\,(\ref{Eq:NS}) subject to the
boundary conditions Eqs.\,(\ref{Eq:BCtop}-\ref{Eq:BCbot}) is given
by
\begin{eqnarray}
u_{x}(z) = \frac{U}{A^2+B^2} \Big\langle \textrm{exp}(+Kz) \times
\big\{ [L_2 K (A+B) + A]\,\textrm{sin}(\omega t + Kz) \nonumber \\
+ [L_2 K (A-B) - B]\,\textrm{cos}(\omega t + Kz) \big\} \nonumber \\
+ \textrm{exp}(-Kz) \times \big\{ [L_2 K
(A+B)-A]\,\textrm{sin}(\omega t - Kz) \nonumber \\ + [L_2 K (A-B) +
B]\,\textrm{cos}(\omega t - Kz) \big\} \Big\rangle,
\label{Eq:Stokes}
\end{eqnarray}
where
\begin{equation}
A = A^{+} - A^{-} \,\,\textrm{and}\,\, B = B^{+} + B^{-},
\label{Eq:AB}
\end{equation}
which in turn are defined as follows:
\begin{eqnarray}
A^{\pm} = \textrm{exp}(\pm Kh)\,\big\{\big[1 \pm
(L_1+L_2)K\big]\,\textrm{cos}(Kh) \nonumber \\ - \big[(L_1+L_2)K \pm
2 L_1 L_2 K^2\big]\,\textrm{sin}(Kh) \big\}, \label{Eq:Adef}
\end{eqnarray}
\begin{eqnarray}
B^{\pm} = \textrm{exp}(\pm Kh)\,\big\{\big[1 \pm
(L_1+L_2)K\big]\,\textrm{sin}(Kh) \nonumber \\ + \big[(L_1+L_2)K \pm
2 L_1 L_2 K^2\big]\,\textrm{cos}(Kh) \big\}, \label{Eq:Bdef}
\end{eqnarray}
and $K = \sqrt{\omega\rho/2\mu}$. In Sect.~\ref{subsec:Oscillatory},
the velocity profiles obtained from MD simulations will be fitted to
Eq.\,(\ref{Eq:Stokes}) with the parameters $L_{1}$ and $L_{2}$.

\subsection{Steady shear flows}
\label{subsec:Steady}

The simulations were first performed at steady-state flow conditions
when the upper wall was translated with a constant velocity, while
the lower wall always remained stationary.   The upper wall velocity
was varied in the range $0.025\,\sigma/\tau\leqslant U\leqslant
6.5\,\sigma/\tau$.   The lower limit was chosen to reduce the
averaging time due to thermal fluctuations, while the upper limit
was set to avoid the nonlinear slip regime at very high shear rates
when the slip velocity becomes much larger than the fluid thermal
velocity~\cite*[]{Niavarani10}.   In the present study, the maximum
slip velocity and shear rate in steady shear flows are about
$1.59\,\sigma/\tau$ and $0.14\,\tau^{-1}$, respectively, which
provide an upper estimate of the Reynolds number $Re\approx29.5$.
It was previously shown for slip flows over periodically corrugated
surfaces that the inertia term in the Navier-Stokes equation
produces a noticeable difference in the slip length at higher
Reynolds numbers of about $130$ \cite*[]{Niavarani08}.

\begin{figure}[t]
\includegraphics[width=12.0cm,angle=0]{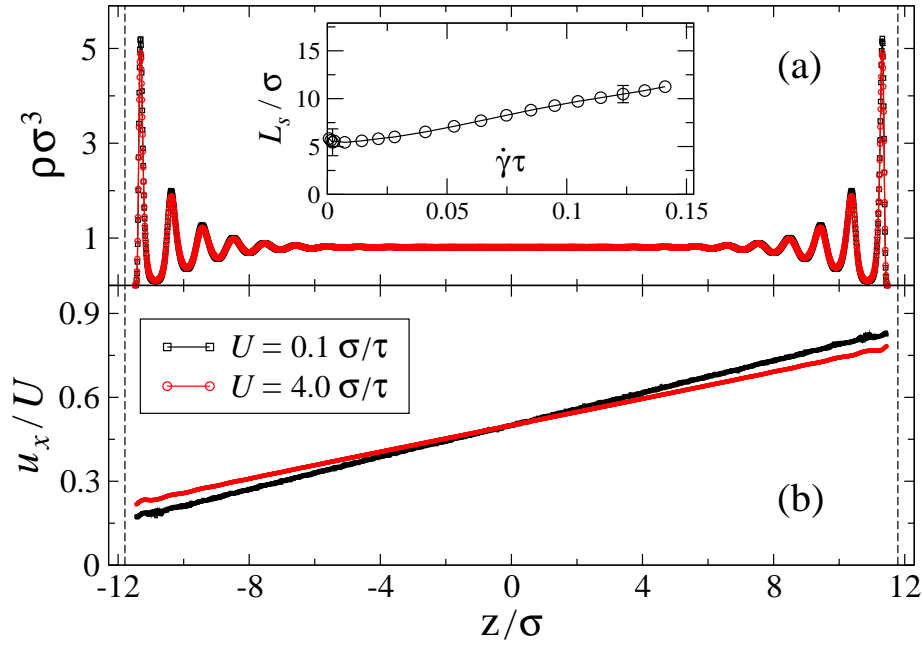}
\caption{(Color online) Ensemble-averaged (a) density and (b)
velocity profiles for the indicated upper wall speeds and
steady-state flow conditions.  The vertical dashed lines at $z=\pm
11.79\,\sigma$ indicate the reference planes for computing the slip
length. The vertical axes at $z=\pm 12.29\,\sigma$ coincide with the
location of the fcc lattice planes which are in contact with the
fluid phase.  The inset shows the slip length as a function of shear
rate when $\omega=0$. } \label{Fig:steady_shear}
\end{figure}

The representative velocity and density profiles for the upper wall
speeds $U=0.1\,\sigma/\tau$ and $U=4.0\,\sigma/\tau$ are plotted in
Fig.\,\ref{Fig:steady_shear}.  As expected, the fluid density
profiles exhibit pronounced oscillations near solid walls that
gradually decay to the uniform bulk value.  The magnitude of the
first density peak defines the contact density $\rho_c$.  Notice
that the amplitude of the density oscillations is slightly reduced
at the higher upper wall speed $U=4.0\,\sigma/\tau$.   The
corresponding velocity profiles, normalized by the upper wall speed,
are linear throughout the channel and are characterized by the
finite slip velocity at both walls.   As clearly observed in
Fig.\,\ref{Fig:steady_shear}\,(b), the relative slip velocity is
larger at the higher upper wall speed.   Also, it was shown
previously for a similar MD setup, that the fluid temperature near
the interfaces increases by about $10\%$ at high shear rates
\cite*[]{Priezjev07,PriezjevJCP}.   The correlation between the
contact density and fluid temperature in the first layer as a
function of the slip velocity was recently reported for polymeric
fluids in steady shear flows \cite*[]{Priezjev12}.

For steady-state flows, the slip length was estimated from the
linear extrapolation of the the velocity profiles to $u_x(z)=0$
below the lower wall and to $u_x(z)=U$ above the upper wall, and
then the two values were averaged.   The variation of the slip
length as a function of shear rate is presented in the inset of
Fig.\,\ref{Fig:steady_shear}\,(a).   In agreement with the results
of previous studies, where the behavior of the slip length was
investigated in a wide range of shear rates and wall-fluid
interaction energies \cite*[]{Priezjev07,PriezjevJCP}, the slip
length increases almost linearly with shear rate when
$\varepsilon_{\rm wf}=\varepsilon$.   It is expected, however, that
when the wall-fluid interaction energy is reduced
\cite*[]{Priezjev07}, then the magnitude of the slip length
increases and its shear rate dependence can be well fitted by the
power-law function proposed by \cite{Nature97}.   In the next
section, the boundary conditions and fluid structure computed in
steady shear flows will be compared with the results obtained for
time-periodic flows.

\subsection{Oscillatory Couette flows}
\label{subsec:Oscillatory}

We next consider oscillatory flows driven by the upper wall,
$u_x^{w}(t) = U \textrm{sin}(\omega t)$, with frequencies
$\omega\tau=10^{-1}, 10^{-2}, 10^{-3}$, and $10^{-4}$.   The
corresponding period and amplitude of oscillations, the Stokes
boundary layer thickness, as well as the upper estimate of the
Reynolds numbers are given in Table\,\ref{Tab:table_I}.   For each
frequency, the amplitude of the velocity oscillation $U$ was chosen
such that the fluid slip velocity at the upper wall was always less
than $1.5\,\sigma/\tau$.   As can be seen from
Table\,\ref{Tab:table_I}, the thickness of the Stokes boundary layer
is smaller than the channel width at higher frequencies
$\omega\tau=10^{-1}$ and $10^{-2}$.   Nevertheless, the upper
estimate of the Reynolds number, based either on the channel width
or the Stokes layer thickness, is about $27.6$, which is indicative
of laminar flow conditions.   In the present study, the smallest
amplitude of the upper wall velocity was set $U=0.25\,\sigma/\tau$
in order to compute accurately the velocity profiles without
excessive computational efforts.

\begin{figure}[t]
\includegraphics[width=12.0cm,angle=0]{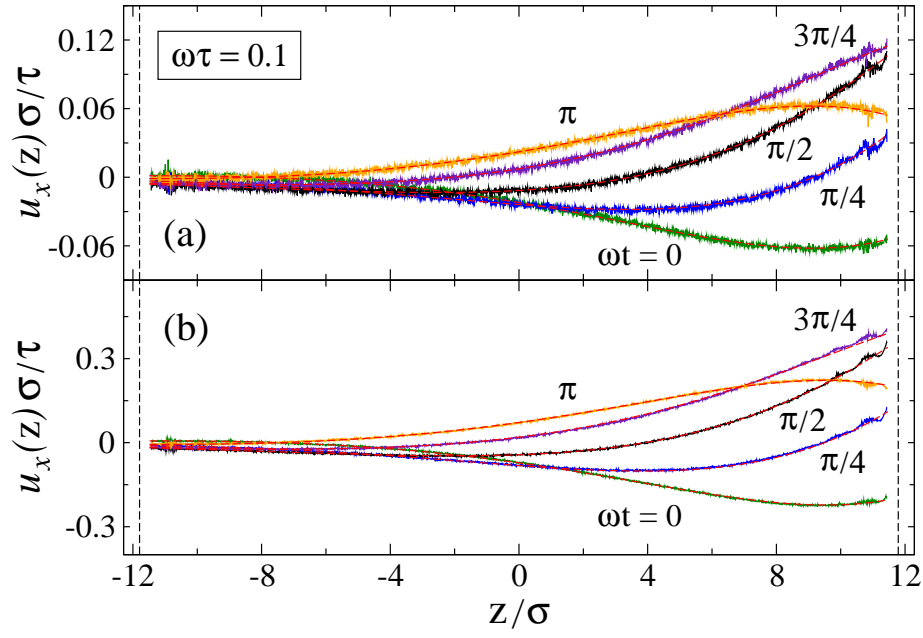}
\caption{(Color online) Averaged velocity profiles for oscillatory
flows with amplitudes (a) $U=0.25\,\sigma/\tau$ and (b)
$U=1.0\,\sigma/\tau$ and frequency $\omega\tau=0.1$ at times $\omega
t = 0, \pi/4, \pi/2, 3\,\pi/4,\,\text{and}\,\pi$.   The red dashed
curves are the least square fits of Eq.\,(\ref{Eq:Stokes}) to the MD
data.   The vertical dashed lines at $z=\pm 11.79\,\sigma$ indicate
the reference planes for computing the slip length and shear rate
from the best fits to Eq.\,(\ref{Eq:Stokes}). }
\label{Fig:velo_omg_0.1}
\end{figure}

\begin{figure}[t]
\includegraphics[width=12.0cm,angle=0]{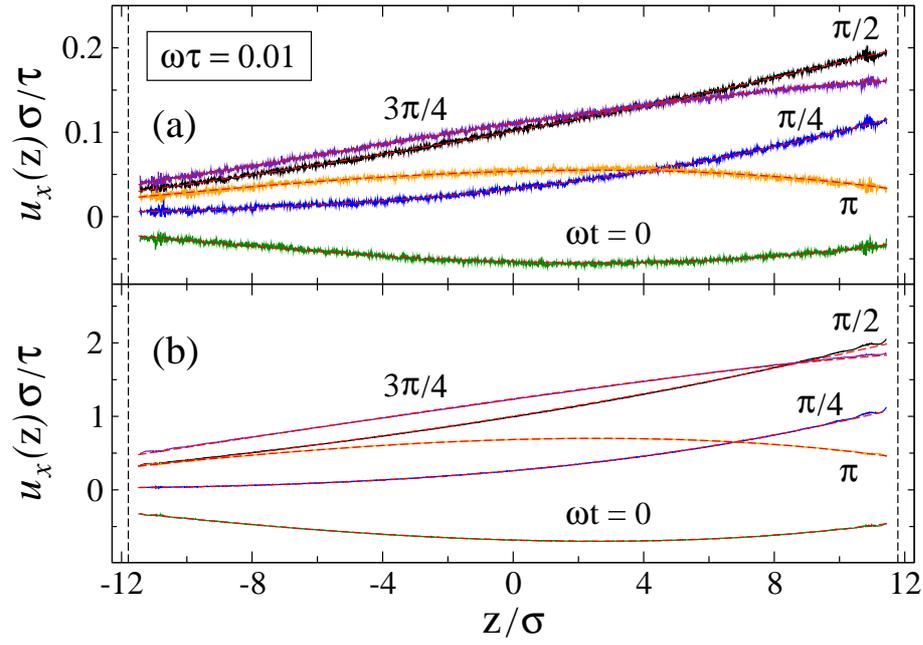}
\caption{(Color online) Velocity profiles for time-periodic flows
with amplitudes (a) $U=0.25\,\sigma/\tau$ and (b)
$U=3.0\,\sigma/\tau$ and frequency $\omega\tau=0.01$ at times
$\omega t = 0, \pi/4, \pi/2, 3\,\pi/4,\,\text{and}\,\pi$.   The red
dashed curves are the best fits of the MD data using
Eq.\,(\ref{Eq:Stokes}).  The vertical dashed lines at $z=\pm
11.79\,\sigma$ denote the location of liquid-solid interfaces.  }
\label{Fig:velo_omg_0.01}
\end{figure}

\begin{figure}[t]
\includegraphics[width=12.0cm,angle=0]{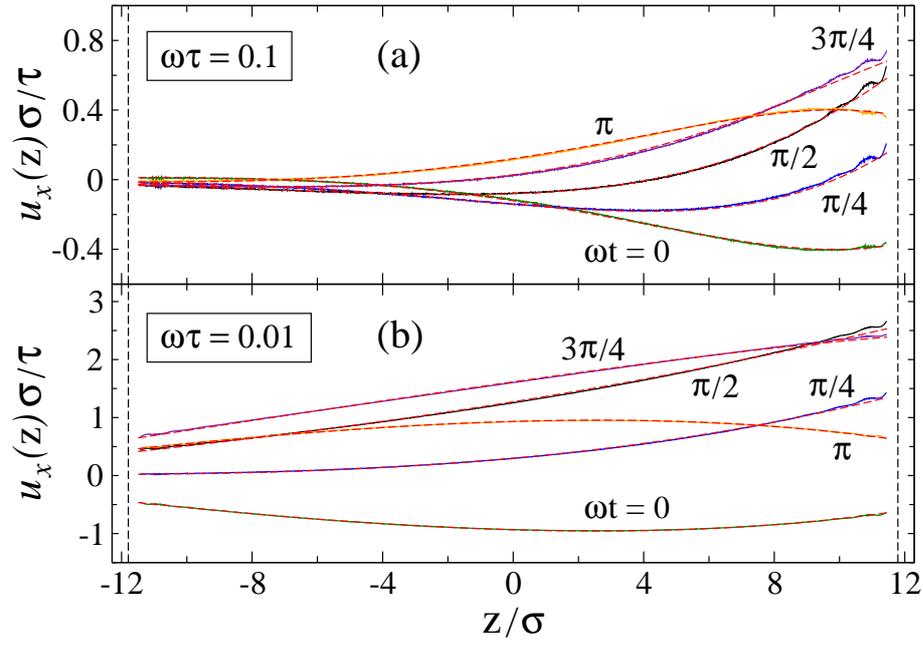}
\caption{(Color online) Averaged velocity profiles for (a)
$\omega\tau=0.1$ and $U=2.0\,\sigma/\tau$ and (b) $\omega\tau=0.01$
and $U=4.0\,\sigma/\tau$ at times $\omega t = 0, \pi/4, \pi/2,
3\,\pi/4,\,\text{and}\,\pi$.  The dashed curves are the best fits of
the MD data using Eq.\,(\ref{Eq:Stokes}).  The corresponding slip
lengths and shear rate magnitudes at the oscillating upper wall are
listed in Table\,\ref{Tab:table_II}.}
\label{Fig:velo_omg_0.1_0.01_highU}
\end{figure}

\begin{figure}[t]
\includegraphics[width=12.0cm,angle=0]{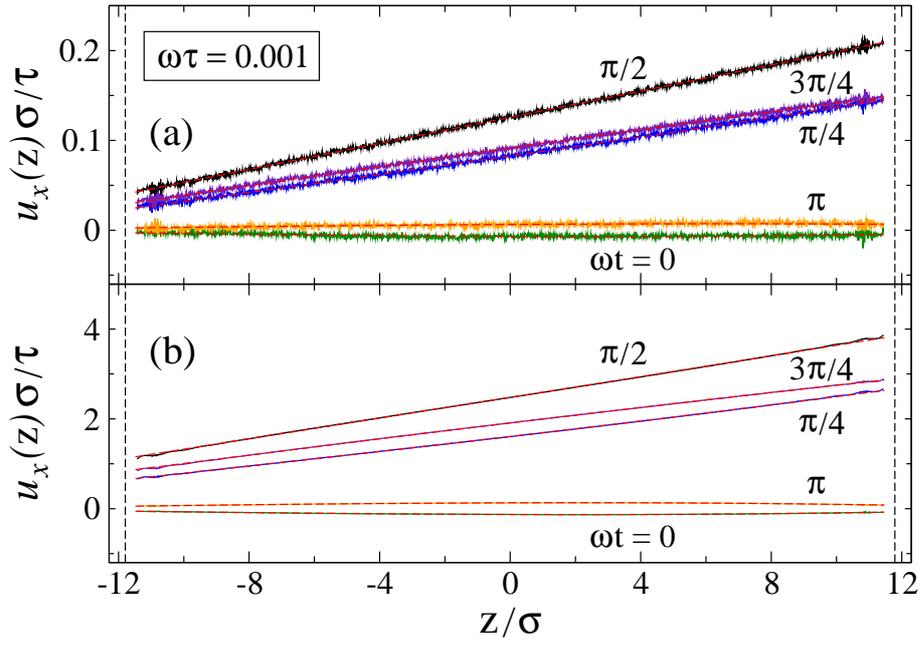}
\caption{(Color online) Velocity profiles for oscillatory flows with
amplitudes (a) $U=0.25\,\sigma/\tau$ and (b) $U=5.0\,\sigma/\tau$
and frequency $\omega\tau=0.001$ at times $\omega t = 0, \pi/4,
\pi/2, 3\,\pi/4,\,\text{and}\,\pi$.  The dashed curves represent the
best fits of the MD data using Eq.\,(\ref{Eq:Stokes}).   The
vertical axes at $z=\pm 12.29\,\sigma$ coincide with the location of
the fcc lattice planes.  } \label{Fig:velo_omg_0.001}
\end{figure}

\begin{figure}[t]
\includegraphics[width=12.0cm,angle=0]{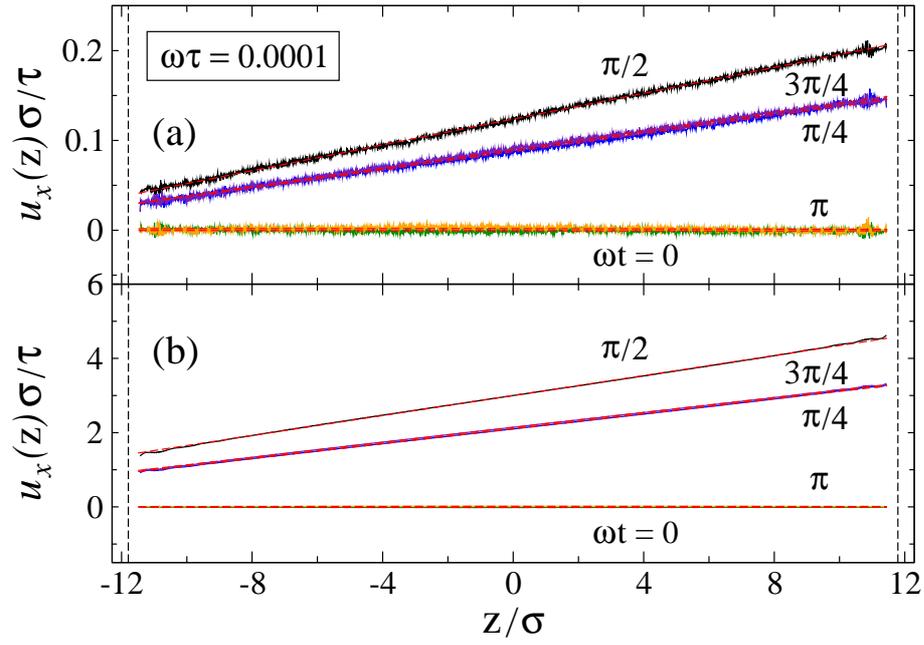}
\caption{(Color online) Velocity profiles for time-periodic flows
with amplitudes (a) $U=0.25\,\sigma/\tau$ and (b)
$U=6.0\,\sigma/\tau$ and frequency $\omega\tau=0.0001$ at times
$\omega t = 0, \pi/4, \pi/2, 3\,\pi/4,\,\text{and}\,\pi$.   The red
dashed curves indicate the best fits of the MD data using
Eq.\,(\ref{Eq:Stokes}).  } \label{Fig:velo_omg_0.0001}
\end{figure}

Examples of the velocity profiles for different frequencies $\omega$
and amplitudes $U$ are presented in Figs.\,\ref{Fig:velo_omg_0.1},
\ref{Fig:velo_omg_0.01}, \ref{Fig:velo_omg_0.1_0.01_highU},
\ref{Fig:velo_omg_0.001}, and \ref{Fig:velo_omg_0.0001}.   The MD
data were averaged over about $20$ periods at the lowest frequency
$\omega\tau=10^{-4}$ and over $2\times10^4$ periods at the highest
frequency $\omega\tau=10^{-1}$. In all figures, the red curves
represent the least square fits of the MD data to
Eq.\,(\ref{Eq:Stokes}) with the parameters $L_{1}$ and $L_{2}$. The
shear rate at the upper and lower walls was then computed by taking
the derivative of the best fit function $\partial u_x/\partial z$ at
$z=\pm 11.79\,\sigma$.   We found that the MD velocity profiles are
well described by the continuum solution Eq.\,(\ref{Eq:Stokes}),
except in the interfacial regions of about $2\,\sigma$ at high shear
rates.   The small discrepancy observed between the MD and continuum
results may originate from the inertial effects and/or the fluid
temperature increase near the walls at high shear rates.

At the highest frequency $\omega\tau=0.1$, the Stokes layer
thickness is nearly three times smaller than the channel width (see
Table\,\ref{Tab:table_I}), and, therefore, the velocity profiles
near the stationary lower wall are not significantly affected by the
moving upper wall, and, as a result, the interfacial shear rate at
the lower wall remains relatively low (see
Fig.\,\ref{Fig:velo_omg_0.1}).   When $\omega\tau=0.01$ in
Fig.\,\ref{Fig:velo_omg_0.01}, the Stokes layer thickness is
approximately equal to the channel width, and the slip velocities
and shear rates at the upper and lower walls become comparable.
Furthermore, at lower frequencies, $\omega\tau=10^{-3}$ and
$10^{-4}$, the flows appear to be quasi-steady and the velocity
profiles are nearly linear throughout the channel (see
Figs.\,\ref{Fig:velo_omg_0.001} and \ref{Fig:velo_omg_0.0001}).   At
the lowest frequency $\omega\tau=10^{-4}$, the velocity profiles are
almost indistinguishable when the magnitude of the upper wall
velocity is the same (e.g., when $\omega t =
\pi/4~\text{and}~3\,\pi/4$ in Fig.\,\ref{Fig:velo_omg_0.0001}).
Notice also that in all cases except $\omega\tau=10^{-4}$, the
velocity profiles at times $\omega t = 0$ and $\pi$ are symmetrical
to each other with respect to the line $u_x=0$ and the fluid slip
velocity is not zero, indicating that the fluid and the upper wall
oscillate with the same frequency but with a finite phase
difference, which is in agreement with the MD results by
\cite{Mohseni12}.

\begin{figure}[t]
\includegraphics[width=12.0cm,angle=0]{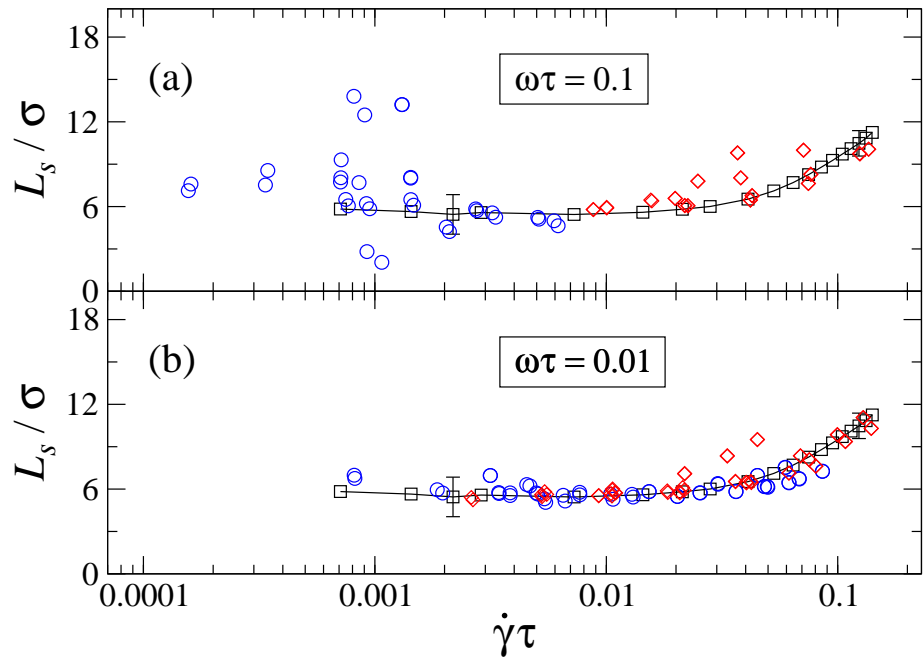}
\caption{(Color online)  The slip length $L_s/\sigma$ as a function
of shear rate for (a) $\omega\tau=0.1$ and (b) $\omega\tau=0.01$.
The slip length and the interfacial shear rate are evaluated at the
stationary lower wall ($\circ$) and at the oscillating upper wall
($\diamond$).  The data for the largest amplitudes $U$ are presented
in Table\,\ref{Tab:table_II}.   The data for steady shear flows
($\square$) are the same as in the inset of
Fig.\,\ref{Fig:steady_shear}\,(a).   The black curves are guides to
the eye.  } \label{Fig:shear_ls_omg_0.1_0.01}
\end{figure}

\begin{figure}[t]
\includegraphics[width=12.0cm,angle=0]{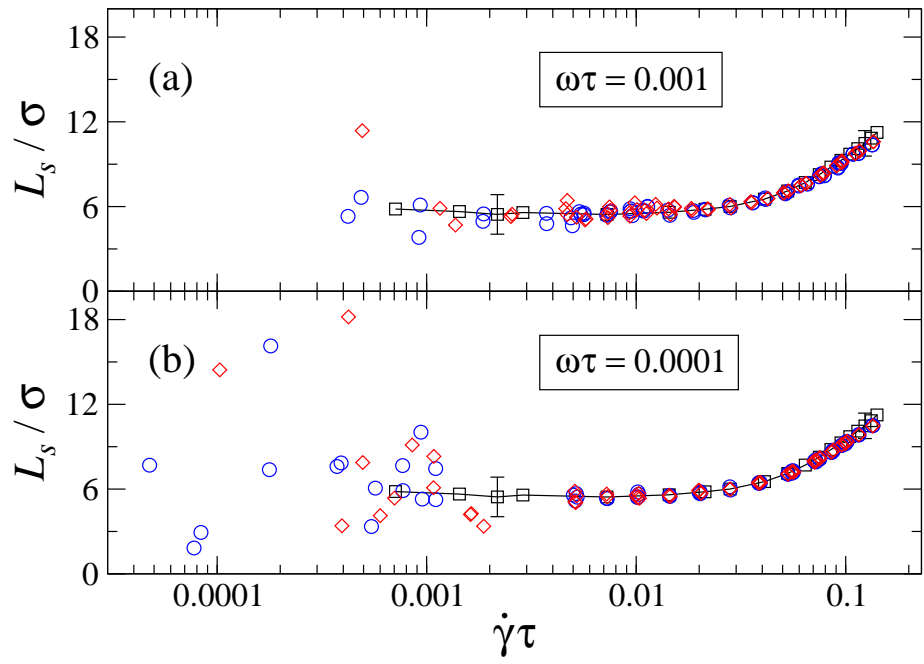}
\caption{(Color online)  Variation of the slip length as a function
of shear rate for (a) $\omega\tau=0.001$ and (b)
$\omega\tau=0.0001$. The slip length and shear rate are computed at
the stationary lower wall ($\circ$) and at the oscillating upper
wall ($\diamond$).   The data ($\square$) are the same as in the
inset of Fig.\,\ref{Fig:steady_shear}\,(a).  }
\label{Fig:shear_ls_omg_0.001_0.0001}
\end{figure}

The continuum solution for oscillatory slip flows
Eq.\,(\ref{Eq:Stokes}) was derived assuming constant slip lengths at
the upper and lower walls.   However, when analyzing the velocity
profiles computed from MD simulations, we noticed that the fitting
parameters $L_{1}$ and $L_{2}$ in Eq.\,(\ref{Eq:Stokes}) depend on
the interfacial shear rate.   The variation of the slip length as a
function of the local shear rate computed at the upper and lower
walls is plotted in Figs.\,\ref{Fig:shear_ls_omg_0.1_0.01} and
\ref{Fig:shear_ls_omg_0.001_0.0001} for different oscillation
frequencies $\omega$, amplitudes $U$, and times $\omega t_n$.   For
comparison, the data for steady shear flows are also presented in
Figs.\,\ref{Fig:shear_ls_omg_0.1_0.01} and
\ref{Fig:shear_ls_omg_0.001_0.0001} on the log-linear scale to
emphasize the low shear rate region.   In all cases, the slip length
for both oscillatory and steady-state flows is nearly constant at
low shear rates $\dot{\gamma}\tau \lesssim 0.01$ and it increases
linearly (see inset in Fig.\,\ref{Fig:steady_shear}) at higher shear
rates.

The deviation from the steady-state results in
Fig.\,\ref{Fig:shear_ls_omg_0.1_0.01} is most pronounced at the
highest frequency $\omega\tau=0.1$ and the largest amplitude
$U=2.0\,\sigma/\tau$, when the magnitude of the wall acceleration is
maximum, i.e., when $\omega t=0$ and $\pi$ [see
Table\,\ref{Tab:table_II} and
Fig.\,\ref{Fig:velo_omg_0.1_0.01_highU}\,(a)].   Notice also that
the MD velocity profiles near the upper wall develop pronounced
oscillations at $\omega t=\pi/4$, $\pi/2$, and $3\,\pi/4$ in
Fig.\,\ref{Fig:velo_omg_0.1_0.01_highU}\,(a).   It can be further
observed that, the data in
Fig.\,\ref{Fig:shear_ls_omg_0.1_0.01}\,(a) are scattered at low
shear rates because the velocity profiles near the lower wall are
not significantly affected by the oscillating upper wall at the
highest frequency $\omega\tau=0.1$ (see
Fig.\,\ref{Fig:velo_omg_0.1}), and the statistical errors due to
thermal fluctuations become relatively large.   Similarly, the
averaged fluid velocity is nearly zero when $\omega t=0$ and $\pi$
at the lowest frequency $\omega\tau=10^{-4}$ (see
Fig.\,\ref{Fig:velo_omg_0.0001}), and, as a result, both $L_s$ and
$\dot{\gamma}$ are subject to statistical uncertainty when
$\dot{\gamma}\tau\lesssim0.005$ in
Fig.\,\ref{Fig:shear_ls_omg_0.001_0.0001}\,(b).   Remember that,
during each cycle, the data were averaged for the time interval
$T/100$, and, thus, significantly longer averaging time is required
to resolve accurately the velocity profiles in oscillating flows. It
is expected, however, that with further averaging, the data in
Figs.\,\ref{Fig:shear_ls_omg_0.1_0.01} and
\ref{Fig:shear_ls_omg_0.001_0.0001} for oscillatory flows at low
shear rates will converge to the steady-state results.


In the case of slip flow over a planar, impermeable solid surface,
the friction coefficient that relates the wall shear stress and slip
velocity is equal $k=\mu/L_s$, when the slip length is computed by
linear extrapolation of the velocity profile to zero
velocity~\cite*[]{Willmott07}.   However, at higher frequencies, as
shown in Fig.\,\ref{Fig:shear_ls_omg_0.1_0.01}, the slip lengths
$L_1$ and $L_2$ computed using Eq.\,(\ref{Eq:Stokes}) deviate from
the slip length in steady-state flows, and, therefore, these values
do not provide an accurate estimate of the friction coefficient.
Also, the estimate of the interfacial shear rate and the
corresponding slip length directly from the MD velocity profiles
(e.g., Figs.\,\ref{Fig:velo_omg_0.1}-\ref{Fig:velo_omg_0.0001}) is
not precise because of the slight nonlinearity of the velocity
profiles near interfaces and the ambiguity in choosing the size of
the fitting region.    To avoid the uncertainty associated with
fitting the velocity profiles, the friction coefficient in
oscillatory flows was estimated from the relation
$\sigma_{xz}(t_n)=k\,u_s(t_n)$.    The wall shear stress
$\sigma_{xz}(t_n)$ was computed as a ratio of the total tangential
force between the fluid monomers and wall atoms to the wall area,
and then averaged over the time interval $T/100$.    At the same
time, the slip velocity was calculated from the velocity and density
profiles as follows:
\begin{equation}
u_{s}=\int_{z_1}^{z_2}\!u_x(z)\rho(z)dz \,\Big/
\int_{z_1}^{z_2}\!\rho(z)dz,   \label{Eq:velo_defin_eq}
\end{equation}
where the integrals were taken over the width of the first peaks in
the density profiles.   Naturally, the fluid slip velocity at the
oscillating upper wall is the difference between the velocity of the
adjacent fluid layer and the upper wall speed.

As was shown in Fig.\,\ref{Fig:steady_shear}\,(a), in the presence
of a solid substrate, the fluid monomers tend to form several
distinct layers that are gradually decaying to a uniform bulk
density. In addition to the density layering, the periodic surface
potential typically induces an in-plane order within the adjacent
fluid layers provided that the wall-fluid interaction energy is
sufficiently high \cite*[]{Thompson90}.   The characteristic
signature of such ordering is the appearance of several sharp peaks
in the static structure factor, which is defined as follows:
\begin{equation}
S(\mathbf{k})=\frac{1}{N_{\ell}}\,\,\Big|\sum_{j=1}^{N_{\ell}}
e^{i\,\mathbf{k}\cdot\mathbf{r}_j}\Big|^2,
\label{Eq:structure_factor}
\end{equation}
where the sum is taken over $N_{\ell}$ fluid monomers in the first
layer and $\mathbf{r}_j=(x_j,y_j)$ is the position vector of the
fluid monomer.   These peaks are most pronounced at the first
reciprocal lattice vectors of the underlying substrate
\cite*[]{Thompson90}.    Examples of the averaged structure factor
and its dependence on the slip velocity were previously reported by
\cite{Priezjev07} for a similar computational setup.    More
recently, it was shown for several liquid-on-solid systems that the
friction coefficient in steady flows correlates well with the
product of the normalized peak in the structure factor and the
contact density of the first fluid layer \cite*[]{Priezjev10}.

\begin{figure}[t]
\includegraphics[width=12.0cm,angle=0]{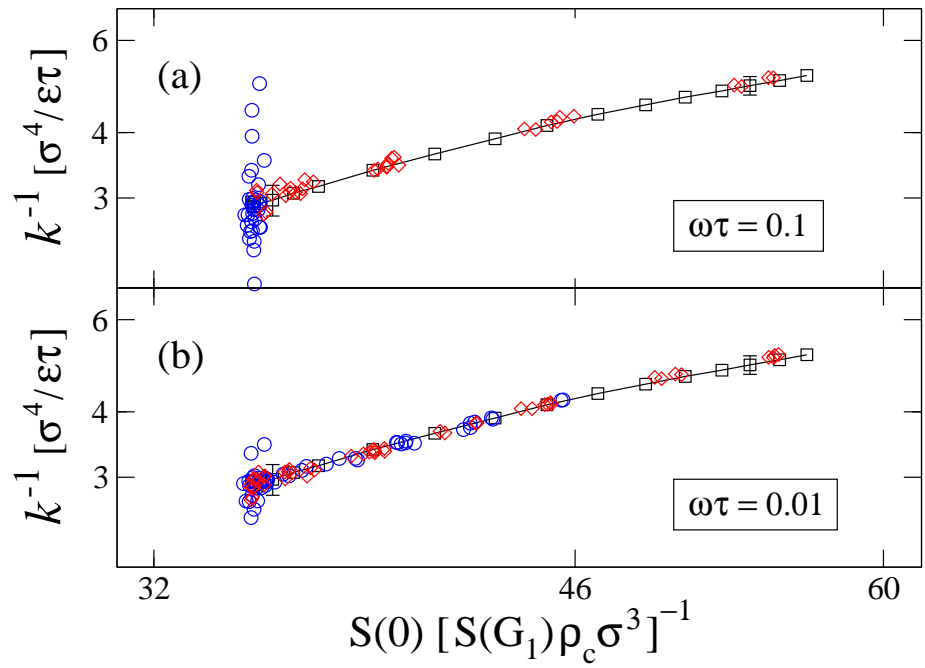}
\caption{(Color online)  Log-log plot of the inverse friction
coefficient as a function of $S(0)/\,[S(\mathbf{G}_1)\,\rho_c]$ for
(a) $\omega\tau=0.1$ and (b) $\omega\tau=0.01$.  The data for
oscillatory flows are denoted by the red diamonds (oscillating upper
wall) and blue circles (stationary lower wall).  The data for steady
shear flows are indicated by the black squares. The black curves are
guides to the eye. }
\label{Fig:inv_fr_vs_S0_div_S7_ro_c_low_omg_0.1_0.01}
\end{figure}

\begin{figure}[t]
\includegraphics[width=12.0cm,angle=0]{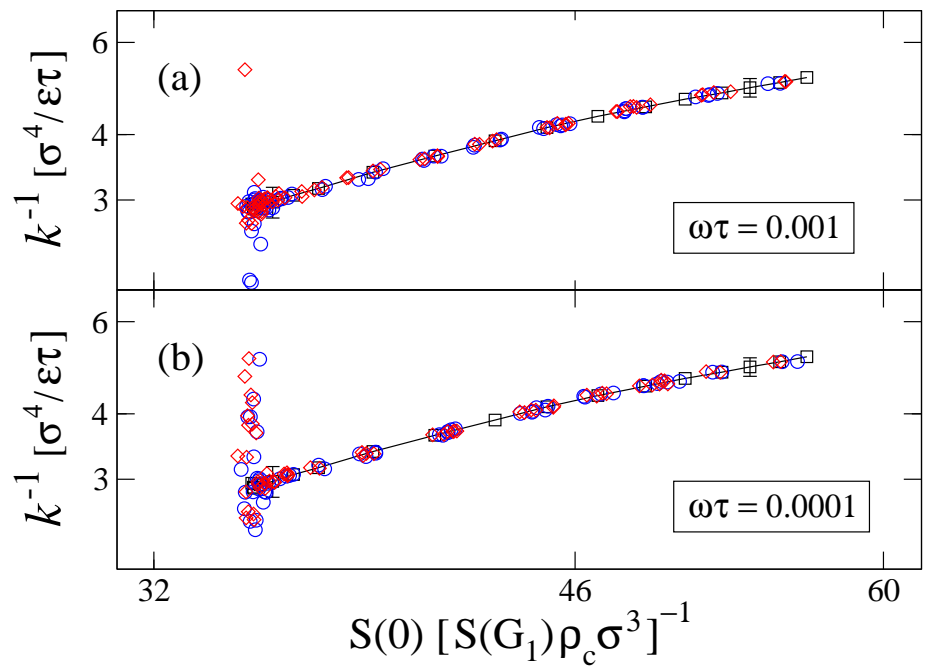}
\caption{(Color online)  The inverse friction coefficient versus
$S(0)/\,[S(\mathbf{G}_1)\,\rho_c]$ for (a) $\omega\tau=0.001$ and
(b) $\omega\tau=0.0001$.    The friction coefficient and fluid
structure are computed at the oscillating upper wall ($\diamond$)
and at the lower stationary wall ($\circ$). The data for steady
shear flows ($\square$) are the same as in
Fig.\,\ref{Fig:inv_fr_vs_S0_div_S7_ro_c_low_omg_0.1_0.01}.  }
\label{Fig:inv_fr_vs_S0_div_S7_ro_c_low_omg_0.001_0.0001}
\end{figure}

In the present study, the friction coefficient in oscillatory and
steady-state flows is plotted in
Figs.\,\ref{Fig:inv_fr_vs_S0_div_S7_ro_c_low_omg_0.1_0.01} and
\ref{Fig:inv_fr_vs_S0_div_S7_ro_c_low_omg_0.001_0.0001} as a
function of the combined variable
$S(0)/\,[S(\mathbf{G}_1)\,\rho_c]$, where
$\mathbf{G}_{1}=(9.04\,\sigma^{-1},0)$ is the first reciprocal
lattice vector in the flow direction.   For both types of flows, the
friction coefficient and the induced fluid structure are reduced at
larger slip velocities.   As is evident, the friction coefficient
extracted from oscillatory flows agrees well with the the
steady-state values, except that the data for periodic flows are
scattered as small slip velocities, which is similar to the rate
dependence of the slip length reported in
Figs.\,\ref{Fig:shear_ls_omg_0.1_0.01} and
\ref{Fig:shear_ls_omg_0.001_0.0001}.   Interestingly, the agreement
at higher frequencies is much better for the friction coefficient
(shown in Fig.\,\ref{Fig:inv_fr_vs_S0_div_S7_ro_c_low_omg_0.1_0.01})
than for the slip length in Fig.\,\ref{Fig:shear_ls_omg_0.1_0.01}.
Note also that at the highest frequency $\omega\tau=0.1$, the period
of oscillation $T=62.83\,\tau$ is about two orders of magnitude
larger than the typical oscillation time of the LJ monomers; but
nevertheless, the structure factor and the contact density are
nearly the same as in steady-state flows.   These results suggest
that slip boundary conditions for high-frequency oscillatory flows
are more accurately described by the dynamic friction coefficient
rather than the slip length as a function of shear rate.

\section{Conclusions}
\label{sec:Conclusions}

In this paper, we have investigated steady and oscillatory Couette
flows with slip boundary conditions using molecular dynamics
simulations.  In both cases, the laminar flows were induced by the
moving upper wall while the lower wall remained stationary.  The
simulations were performed in a wide range of oscillation
frequencies; namely, when the Stokes boundary layer thickness is
smaller than the channel width at the highest frequency, and, on the
other hand, at lower frequencies that correspond to quasi-steady
flows.   For the chosen parameters, the liquid and solid phases form
incommensurate structures at the interface, which is characterized
by a finite slip length that increases almost linearly with shear
rate.

We found that the velocity profiles computed in MD simulations are
well described by the Stokes flow solution with the slip length as a
fitting parameter that depends on the local shear rate.  The rate
dependence of the slip length obtained in steady-state shear flows
is reproduced in oscillatory flows when the slip length is measured
as a function of the absolute value of the local shear rate.  The MD
data for oscillatory flows at low shear rates are relatively noisy,
which, however, is not surprising given that the velocity profiles
were averaged over a small fraction of the period during each cycle.
For both types of flows, the friction coefficient at the
liquid-solid interface correlates well with the structure factor and
the contact density of the first fluid layer.

\begin{acknowledgements}

Financial support from the National Science Foundation
(CBET-1033662) is gratefully acknowledged. Computational work in
support of this research was performed at Michigan State
University's High Performance Computing Facility.

\end{acknowledgements}

\newpage

\begin{table}
\caption{The oscillation frequency $\omega$ (in units $\tau^{-1}$),
the oscillation period $T=2\,\pi/\omega$ (in units $\tau$), the
maximum amplitude of the upper wall velocity (in units
$\sigma/\tau$), the upper estimate of the Reynolds number
$Re_{max}=\triangle Uh\rho/\mu$, the Stokes boundary layer thickness
$\delta=\sqrt{2\,\mu/\rho\,\omega}$ (in units $\sigma$), and the
corresponding Reynolds number $Re^{\delta}_{max}=\triangle U
\delta\rho/\mu$, when $\delta < h = 23.58\,\sigma$.  In the
definition of the Reynolds numbers, $\triangle U$ is the maximum
variation of the tangential fluid velocity component across the
channel. } \label{Tab:table_I}
\begin{tabular}{llllll}
\hline\noalign{\smallskip}
$\omega\,\tau$ & $T/\tau$ & $U_{max}$ & $Re_{max}$ & $\delta/\sigma$ & $Re^{\delta}_{max}$ \\
\noalign{\smallskip}\hline\noalign{\smallskip}
0.1     &  62.83     &  2.0  &  6.5   &  7.29   &  2.0   \\
0.01    &  628.32    &  4.0  &  18.8  &  23.04  &  18.4  \\
0.001   &  6283.19   &  6.0  &  27.6  &  72.86  &  $-$   \\
0.0001  &  62831.85  &  6.0  &  27.6  &  230.4  &  $-$   \\
\noalign{\smallskip}\hline
\end{tabular}
\end{table}

\begin{table}
\caption{The instantaneous slip lengths and shear rate magnitudes at
the oscillating upper wall obtained from the best fit of the MD
velocity profiles in Fig.\,\ref{Fig:velo_omg_0.1_0.01_highU} using
Eq.\,(\ref{Eq:Stokes}).   The amplitude of the upper wall velocity
is $U=2.0\,\sigma/\tau$ for $\omega\tau=0.1$ and
$U=4.0\,\sigma/\tau$ for $\omega\tau=0.01$.   The same data as in
Fig.\,\ref{Fig:shear_ls_omg_0.1_0.01}. } \label{Tab:table_II}
\begin{tabular}{lllllllll}
\hline\noalign{\smallskip}
$\omega t$ & $0$ & $\pi/4$ & $\pi/2$ & $3\pi/4$ & $~\pi$ & $5\pi/4$ & $3\pi/2$ & $7\pi/4$ \\
\noalign{\smallskip}\hline\noalign{\smallskip}
$\omega\tau=0.1$ & $$ & $$ & $$ & $$ & $$ & $$ & $$ & $$ \\
$\dot{\gamma}\tau$ & $0.037$ & $0.124$ & $0.136$ & $0.071$ & $0.037$ & $0.125$ & $0.136$ & $0.071$ \\
$L_s/\sigma$ &       $9.8$ &   $9.8$ &   $10.1$ &  $10.0$ &  $9.8$ &   $9.7$ &   $10.1$ &  $10.0$ \\
\noalign{\smallskip}\hline\noalign{\smallskip}
$\omega\tau=0.01$ & $$ & $$ & $$ & $$ & $$ & $$ & $$ & $$ \\
$\dot{\gamma}\tau$ & $0.080$ & $0.140$ & $0.129$ & $0.045$ & $0.080$ & $0.140$ & $0.129$ & $0.045$ \\
$L_s/\sigma$ &       $7.6$ &   $10.3$ &  $11.1$ &  $9.5$ &   $7.6$ &   $10.3$ &  $11.1$ &  $9.5$ \\
\noalign{\smallskip}\hline
\end{tabular}
\end{table}

\newpage


\end{document}